\documentclass{Interspeech}

\interspeechcameraready

\title{VibE-SVC: Vibrato Extraction with High-frequency F0 Contour for Singing Voice Conversion}

\author[]{Joon-Seung}{Choi}
\author[]{Dong-Min}{Byun}
\author[]{Hyung-Seok}{Oh}
\author[affiliation={{\dagger}}]{Seong-Whan}{Lee}

\affiliation[nocounter]{Department of Artificial Intelligence}{Korea University}{Seoul, Korea}
\email{js\_choi@korea.ac.kr, dm\_byun@korea.ac.kr, hs\_oh@korea.ac.kr, sw.lee@korea.ac.kr\thanks{$^\dagger$Corresponding author}}


\usepackage{comment, subcaption, graphicx, multirow}
\usepackage{hyperref}
\begin{document}

\maketitle

\begin{abstract}

Controlling singing style is crucial for achieving an expressive and natural singing voice. Among the various style factors, vibrato plays a key role in conveying emotions and enhancing musical depth. However, modeling vibrato remains challenging due to its dynamic nature, making it difficult to control in singing voice conversion. To address this, we propose VibE-SVC, a controllable singing voice conversion model that explicitly extracts and manipulates vibrato using discrete wavelet transform. Unlike previous methods that model vibrato implicitly, our approach decomposes the F0 contour into frequency components, enabling precise transfer. This allows vibrato control for enhanced flexibility. Experimental results show that VibE-SVC effectively transforms singing styles while preserving speaker similarity. Both subjective and objective evaluations confirm high-quality conversion. 
\end{abstract}

\section{Introduction}

Singing voice conversion (SVC) is a technique that converts a source singer voice into a target singer voice while retaining the source lyrics, melody, and styles. Recently, many singing voice models have been developed using various generative models \cite{USVC, VISinger, diffsinger, diffsvc, byun2024midi}. Although SVC has improved significantly, several challenges remain. One of the most important challenges is effectively handling pitch. Since the singing voice is more expressive than speech, accurate pitch modeling is essential.

For this reason, several studies \cite{SVCC_T23, Spa-svc} have been proposed to improve pitch-related performance. SVCC-T23 \cite{SVCC_T23} extracts multi-scale F0 as an auxiliary input to better capture pitch variation in singing. SPA-SVC \cite{Spa-svc} introduces a cycle pitch shifting strategy to mitigate voiceless regions and hoarse artifacts caused by narrow pitch range of input dataset. 

Some works \cite{Vibrato_modeling, TCSinger,SinTechSVS} have focused on handling style characteristics of singing, including pitch styles. Vibrato control is achieved in \cite{Vibrato_modeling} by extracting vibrato extent from the power spectrogram of the first-order difference. To synthesize and control singing styles, SinTechSVS \cite{SinTechSVS} proposes style recommender and singing technique local score module. TCSinger \cite{TCSinger} introduces clustering style encoder to capture singing styles. 
Unlike these methods which focus on synthesizing singing styles, other approaches \cite{ PST, gmvae-stc, MSN-STC, svc_refencoder} focus on transferring singing styles. 
Since style features are not explicitly defined, these methods disentangle style information implicitly.

Extracting information via signal decomposition is a key focus in various deep learning studies \cite{prml_video, prml_bci}. Discrete wavelet transform is a method for decomposing an arbitrary signal into functions defined by discretely sampled wavelets. Wavelet transform has been utilized in various methods \cite{wavelet_MCNN, fastspeech2, dwt_singer_identification, fregan2} such as pitch representation \cite{fastspeech2}, singer identification \cite{dwt_singer_identification}, or a downsampling method \cite{fregan2}. 
We assume that while the overall F0 contour remains consistent across singing styles, style-related variations are reflected in the high-frequency contour. To capture these differences, we use DWT to decompose the F0 contour into low- and high-frequency bands. Most existing methods aim to smooth the F0 contour to eliminate minor fluctuations or singing styles using filters such as median filter \cite{periodsinger} or band-pass filter \cite{band_filter}. 
In contrast, our method employs DWT as a pass filter to disentangle and control singing styles.

In this work, we propose VibE-SVC, which disentangles vibrato style from singing voices and enables style transfer using DWT.
By predicting the high-frequency F0 contour, VibE-SVC achieves transfer between straight and vibrato styles. Our approach demonstrates that DWT effectively separates singing styles and allows vibrato extent control without explicitly modeling the vibrato extent feature. Experimental results show that VibE-SVC successfully transfers singing styles. Audio samples are available at \url{https://castlechoi.github.io/VibE-SVC-demo}.

\begin{figure*}
  \centering
  \begin{subfigure}[t]{0.43\textwidth}
      \includegraphics[width=\textwidth]{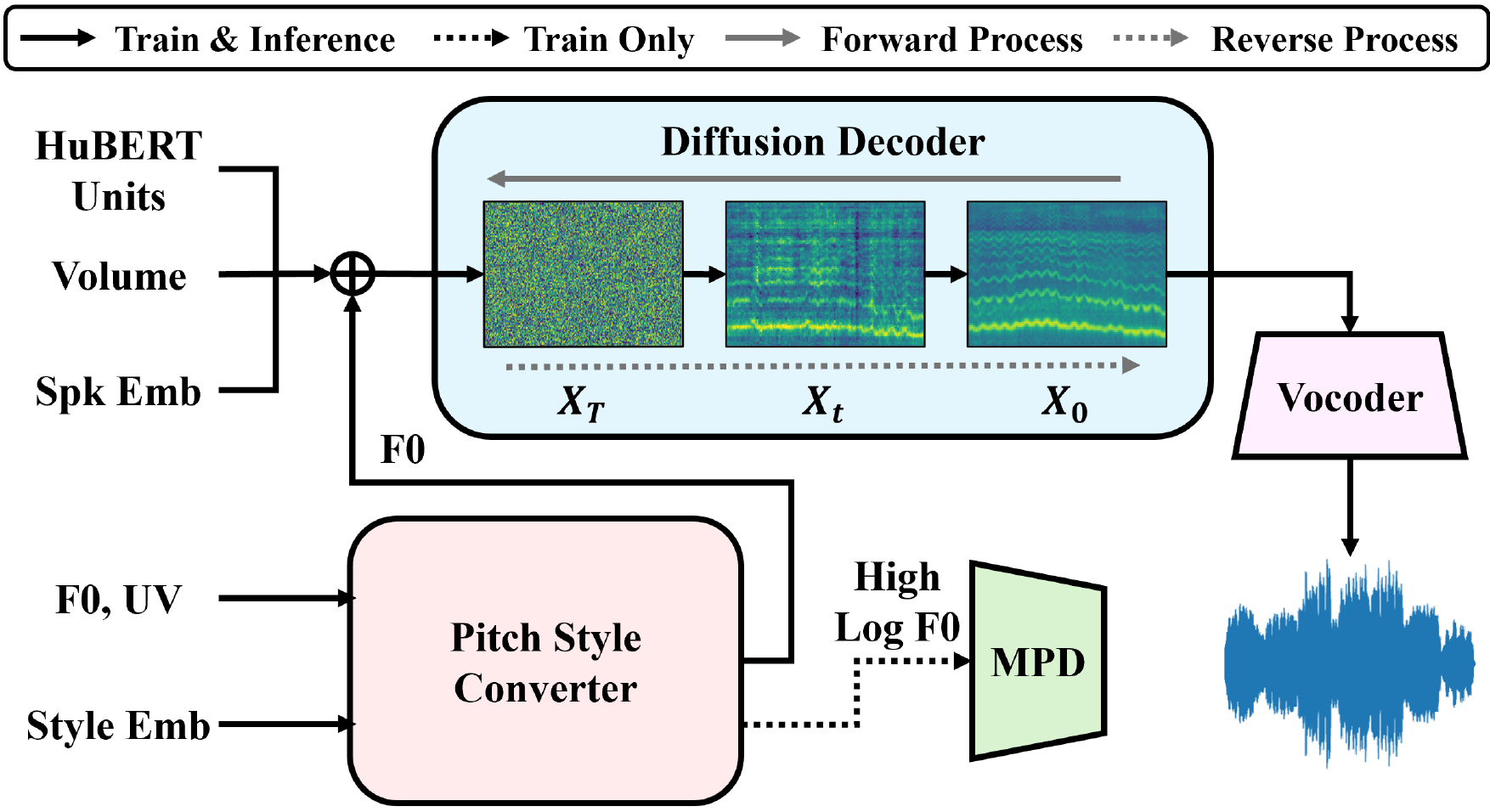}
  \caption{VibE-SVC}
  \label{fig:svc_architecture}
  \end{subfigure}
  \begin{subfigure}[t]{0.55\textwidth}
      \includegraphics[width=\textwidth]{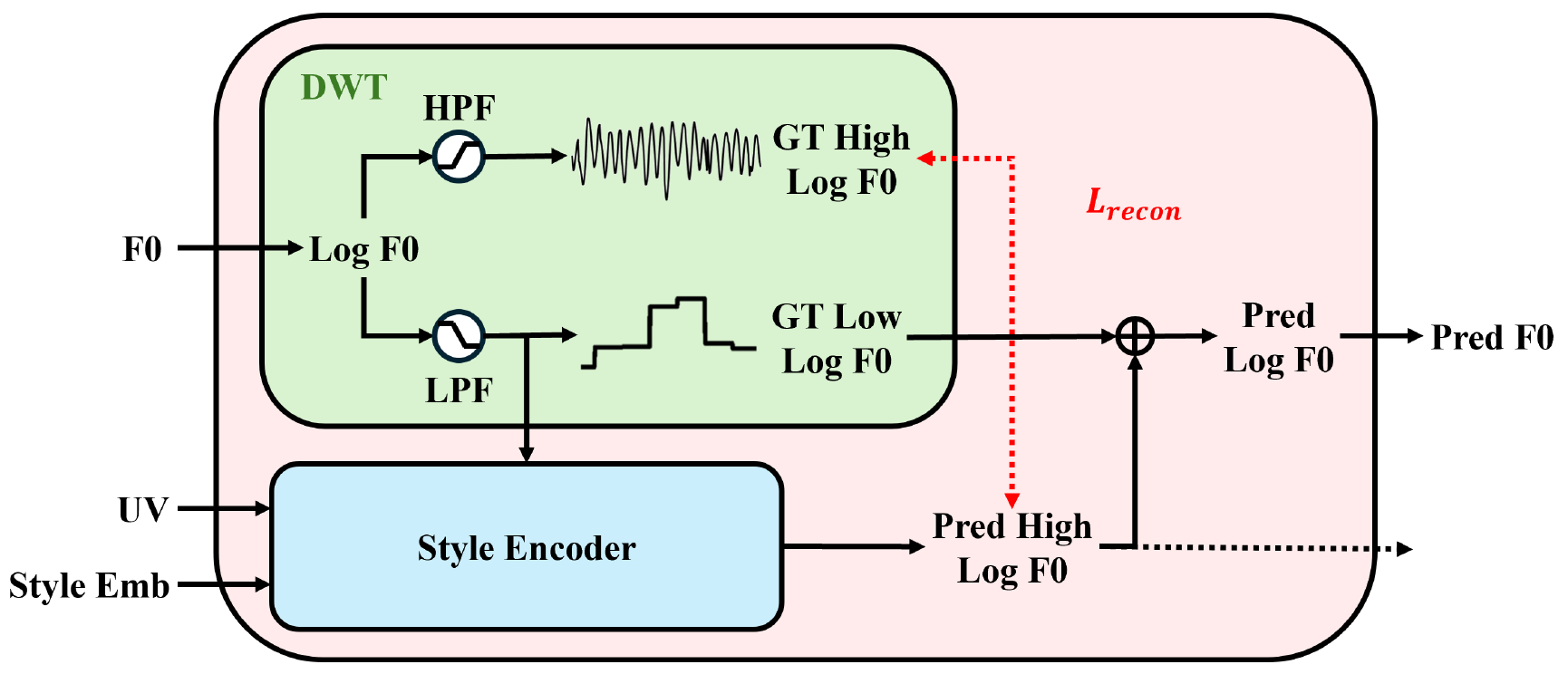}
  \caption{Pitch style converter}
  \label{fig:converter_architecture}
  \end{subfigure}
  \vspace{-0.2cm}
  \caption{The overview of the proposed VibE-SVC model.}
  \vspace{-0.4cm}
  \label{fig:overall_architecture}
\end{figure*}

\section{Method}
We propose a controllable SVC model that enables style conversion between straight and vibrato. To separate style-related variations, we employ a DWT-based method to decompose the F0 contour into low- and high-frequency contours. The high-frequency contour is then predicted by a pitch style converter to perform singing style conversion. An overview of the VibE-SVC framework is shown in Figure \ref{fig:overall_architecture}.

\subsection{Vibrato disentanglement}
\begin{figure}
  \centering
  
  \begin{subfigure}[t]{0.95\linewidth}
  \centering
  \includegraphics[width=\linewidth]{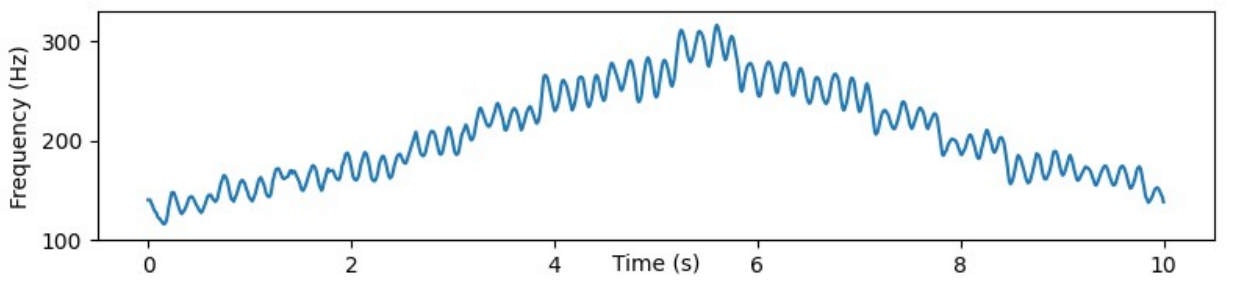}
  \vspace{-0.5cm}
  \caption{Source F0 contour}
  \label{fig:orig_pitch}
  \end{subfigure}
  
  \begin{subfigure}[t]{0.95\linewidth}
  \includegraphics[width=\linewidth]{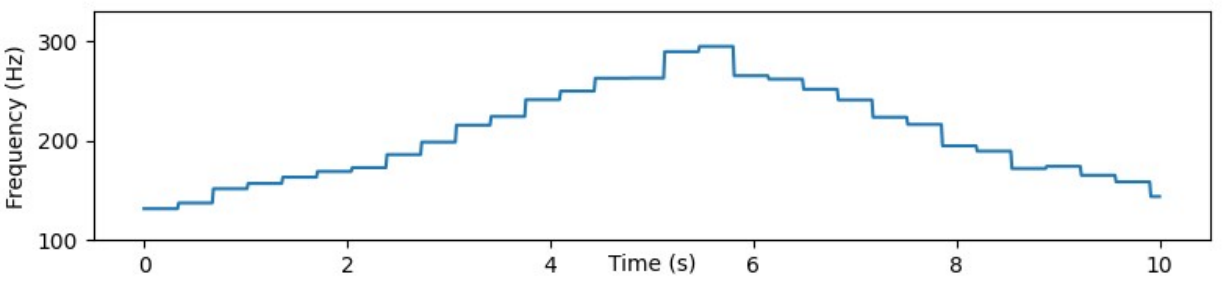}
  \vspace{-0.5cm}
  \caption{Low-frequency F0 contour}
  \label{fig:low_pitch}
  \end{subfigure}
  
  \begin{subfigure}[t]{0.95\linewidth}
  \centering
  \includegraphics[width=\linewidth]{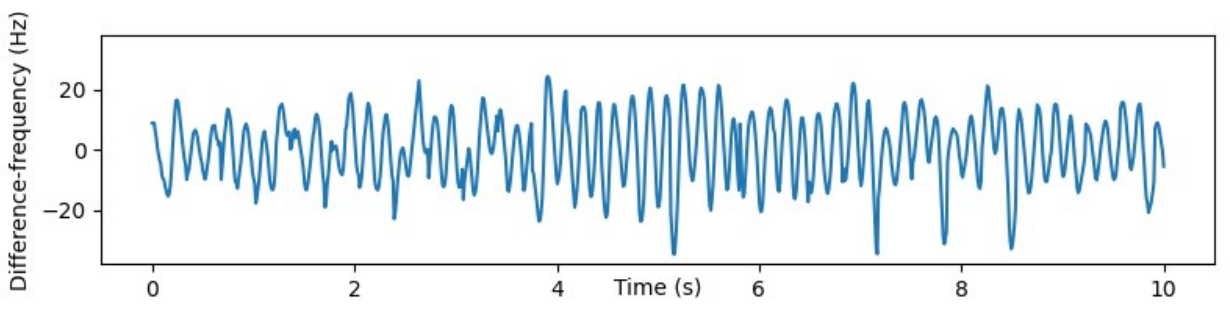}
  \vspace{-0.5cm}
  \caption{High-frequency F0 contour}
  \label{fig:high_pitch}
  \end{subfigure}
  
  \caption{F0 contour disentanglement: (a) Source F0 contour, (b) Reconstructed F0 contour from approximation coefficient, and (c) Reconstructed F0 contour from detail coefficients.}
  \vspace{-0.5cm}
  \label{fig:dwt_analysis}
\end{figure}

We use DWT to decompose a signal into approximation and detail coefficients, corresponding to the low- and high-frequency components. Approximation coefficient $A$ and detail coefficient $D$ are defined as follows:
 \begin{align}
    A_j[k] &= \sum_nx[n]\phi_{j,k}[n], \label{equation:dwt_approx} \\
    D_j[k] &= \sum_nx[n]\psi_{j,k}[n],\label{equation:dwt_detail}
\end{align}
where $x[n]$ denotes source signal. 
$j$, $k$, and $n$ denote the decomposition level of DWT, the position of wavelet function, and the position of signal $x$, respectively. 
$\phi$ and $\psi$ denote the scaling function and the wavelet function, respectively. To reconstruct the low- and high-frequency signals from the corresponding coefficients, we use the inverse discrete wavelet transform (iDWT) as follows:
\begin{align}
x[n] &= \sum_kA_{L}[k]\phi_{L,k}[n]
+ \sum_{j=1}^{L}\sum_kD_j[k]\psi_{j,k}[n], 
\label{equation:iDWT}
\end{align}
where $L$ denotes the lowest frequency DWT level. 
Based on Equation \ref{equation:iDWT}, we reconstruct low- and high-frequency F0 contours from the approximation coefficient and detail coefficients. As shown in Figure \ref{fig:dwt_analysis}, high-frequency F0 contour $x_{high}$ and low-frequency F0 contour $x_{low}$ are defined as follows:
\begin{align}
x_{high}[n] &= \sum_kA_{L}[k]\phi_{L,k}[n], \label{equation:iDWT_high}\\
x_{low}[n] &= \sum_{j=1}^L\sum_kD_j[k]\psi_{j,k}[n].\label{equation:iDWT_low}
\end{align}
Based on Equations \ref{equation:iDWT_high} and \ref{equation:iDWT_low}, the source F0 contour is reconstructed by low- and high-frequency F0 contours as follows:
\begin{align}
x[n] &= x_{low}[n] + x_{high}[n]. \label{equation:iDWT_recon}
\end{align}
We adopt the Daubechies1 (db1) wavelet function \cite{db1} to enhance the extraction of consistent vibrato extent and facilitate model training. In DWT, the db1 wavelet function produces a rectangular function that aligns well with the characteristics of the musical instrument digital interface (MIDI).

\subsection{Pitch style converter}
As shown in Figure \ref{fig:converter_architecture}, we use log F0 contour to normalize the vibrato scale across frequencies, and decompose it into frequency components using DWT. The low-frequency F0 contour, voice flag vector, and style embedding obtained from the style lookup table are concatenated and used as input to the style encoder to predict the high-frequency F0 contour as the target. After prediction, the log F0 contour converted to the target style is reconstructed by Equation \ref{equation:iDWT_recon}. The predicted log F0 contour is then denormalized and fed into the diffusion decoder.

Since the vibrato rate is typically characterized by a frequency range of 5 to 8 Hz \cite{Vibrato_modeling, vibrato_detection}, we adopt a multi-period discriminator (MPD) \cite{hifigan} to capture the periodic characteristics in the high-frequency F0 contour. The predicted and ground-truth high-frequency log F0 contour are used as the input to the MPD.

We train the model using reconstruction loss $L_{recon}$ for log F0 contour, adversarial loss $L_{adv}$ \cite{lsgan},
and feature matching loss $L_{fm}$. The training loss functions are defined as follows: 
\begin{align}
L(G) &= L_{recon}(G)+\mathcal L_{fm}(G)+\mathcal L_{adv}(G), \label{equation:generator_loss}\\
L(D) &= L_{adv}(D), \label{equation:discriminator_loss}
\end{align}
where $G$ denotes the generator and $D$ denotes the discriminator.

\subsection{Overall architecture}
As shown in Figure \ref{fig:overall_architecture}, VibE-SVC consists of two main components: the SVC model and the pitch style converter, which are trained separately. The SVC model generates Mel-spectrograms conditioned on HuBERT \cite{HuBERT} units, speaker embeddings, the F0 contour, and volume. The HuBERT units are interpolated to align with the F0 contour, while the volume is computed as the squared magnitude of the input audio.
The SVC model is optimized with a simple diffusion loss, whereas the pitch style converter is trained according to the method described in Section 2.2.
For training, the source speaker embedding and source F0 contour are used as inputs.
During inference, we provide the target style embedding to predict the style-transferred F0 contour. Then, we provide the target speaker embedding and the predicted F0 contour as input to the SVC model. To adjust the pitch range, we scale the F0 contour by the ratio of the mean F0 values of the source and target speakers. The mean F0 value for each speaker is computed in advance.

\begin{table*}[th]
  \caption{Comparison results of style transfer. The MOS and SMOS are presented with 95\% confidence intervals.}
  \vspace{-0.25cm}
  \label{tab:main_table}
  \centering
  \resizebox{1.00\textwidth}{!}{
  \centering
  \begin{tabular}{l|cc|cc|cc|cc}
    \toprule
    \multicolumn{1}{c|}{} & 
    \multicolumn{4}{c|}{\textbf{Style-Only Conversion}}     & 
    \multicolumn{4}{c}{\textbf{Timbre \& Style Conversion}} \\
    \midrule
    
    \multicolumn{1}{c|}{\textbf{Method}}& 
    \multicolumn{1}{c}{\textbf{MOS}}    &
    \multicolumn{1}{c|}{\textbf{SMOS}}  &
    \multicolumn{1}{c}{\textbf{SECS}}   & 
    \multicolumn{1}{c|}{\textbf{Acc}}   &
    
    \multicolumn{1}{c}{\textbf{MOS}}    &
    \multicolumn{1}{c|}{\textbf{SMOS}}  &
    \multicolumn{1}{c}{\textbf{SECS}}   & 
    \multicolumn{1}{c}{\textbf{Acc}}   \\
    \midrule

    \textbf{GT} 
    &\text{4.223 $\pm$ 0.08} &\text{3.355 $\pm$ 0.09} 
    &\text{-} &\text{-}  
    &\text{4.242 $\pm$ 0.09} &\text{3.549 $\pm$ 0.06} 
    &\text{-} &\text{-}  \\

    \textbf{Vocoded} 
    &\text{4.221 $\pm$ 0.07} &\text{3.340 $\pm$ 0.09} 
    &\text{0.818} &\text{0.975}  
    &\text{4.290 $\pm$ 0.09} &\text{3.535 $\pm$ 0.06} 
    &\text{0.818} &\text{0.975}  \\
    
    \midrule
    
    \textbf{SoVITS w/ Style Emb} 
    &\textbf{4.197 $\pm$ 0.08} &\textbf{2.911 $\pm$ 0.12} 
    &\text{0.800} &\text{0.213}  
    &\text{4.112 $\pm$ 0.11} &\text{3.221 $\pm$ 0.09} 
    &\text{0.768} &\text{0.179}   \\
    
    \textbf{SoVITS w/ Style Emb \& DWT} 
    &\text{4.078 $\pm$ 0.09} &\text{2.872 $\pm$ 0.13} 
    &\text{0.798} &\text{0.525}
    &\text{3.997 $\pm$ 0.12} &\textbf{3.224 $\pm$ 0.10} 
    &\text{0.765} &\text{0.492}   \\
    
    \textbf{SoVITS w/ PST} 
    &\text{4.043 $\pm$ 0.09} &\text{2.897 $\pm$ 0.13} 
    &\text{0.804} &\text{0.425} 
    &\text{4.035 $\pm$ 0.11} &\textbf{3.224 $\pm$ 0.09} 
    &\textbf{0.774} &\text{0.434} \\
    
    \midrule
    
    \textbf{VibE-SVC (Ours)}
    &\text{4.124 $\pm$ 0.08} &\textbf{2.911 $\pm$ 0.13} 
    &\textbf{0.814} &\textbf{0.700}  
    &\textbf{4.125 $\pm$ 0.10} &\text{3.196 $\pm$ 0.09}
    &\textbf{0.774} &\textbf{0.694}\\
    
    \textbf{w/o MPD}   
    &\text{4.016 $\pm$ 0.09} &\text{2.847 $\pm$ 0.13} 
    &\text{0.809} &\text{0.625} 
    & \text{3.997 $\pm$ 0.11} &\text{3.218 $\pm$ 0.09} 
    &\text{0.773} &\text{0.611}  \\
    
    \textbf{w/o DWT}  
    &\text{4.004 $\pm$ 0.10} &\text{2.887 $\pm$ 0.13} 
    &\text{0.800} &\text{0.475}
    &\text{3.971 $\pm$ 0.12} &\text{3.213 $\pm$0.10} 
    &\text{0.769} &\text{0.535} \\
    \bottomrule
  \end{tabular}
  }
  \vspace{-0.4cm}
\end{table*}

\begin{figure}
  \centering
  \begin{subfigure}[t]{0.48\linewidth}
  \includegraphics[width=\linewidth]{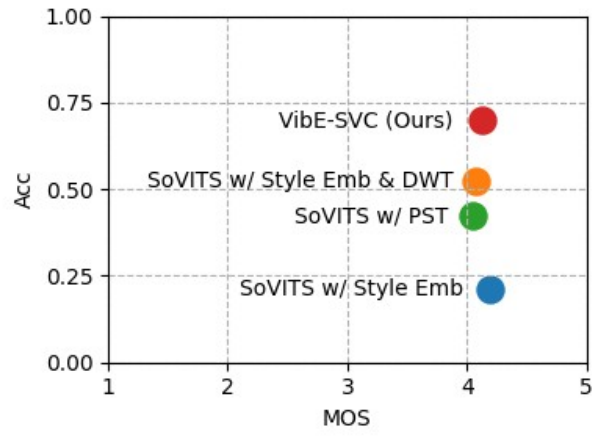}
  \caption{Style-Only conversion}
  \label{fig:style_only_conversion}
  \end{subfigure}
  \begin{subfigure}[t]{0.49\linewidth}
  \includegraphics[width=\linewidth]{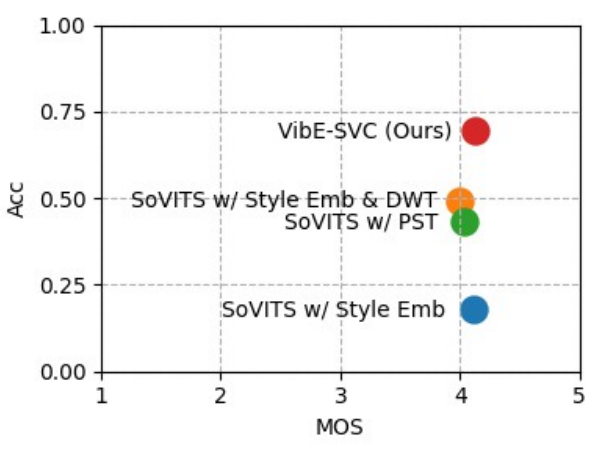}
  \caption{Timbre \& Style conversion}
  \label{fig:timbre_style_conversion}
  \end{subfigure}
  \caption{Correlation between MOS and style accuracy.}
  \label{fig:mos_acc_corr}
  \vspace{-0.4cm}
\end{figure} 

\section{Experiments}

\subsection{Dataset}
We use VocalSet \cite{vocalset}, a singing voice dataset labeled with 17 different styles. The dataset includes 20 singers, consisting of 11 males and 9 females. For our experiments, we select the straight and vibrato styles from the dataset, resulting in a total of 753 samples. All audio recordings are downsampled to 24 kHz. We set the hop size to 256, the FFT size to 1024, and the window size to 1024 for Mel-spectrogram extraction. For training efficiency, each audio clip is segmented into 2-second chunks. For the test set, 4 samples from both straight and vibrato styles are selected per singer, making a total of 160 samples. The remaining data is used for training. The F0 contour and voiced flag vector are extracted using DIO \cite{dio}.

\subsection{Implementation details}
VibE-SVC is based on an open-source SVC project\footnote{\url{https://github.com/svc-develop-team/so-vits-svc}}, which uses a WaveNet \cite{wavenet}-based diffusion decoder with 1000 steps. The diffusion decoder consists of 20 layers in the residual blocks, 512 output channels in the convolutional layers, a hidden dimension of 256, and an encoder hidden layer size of 256. We use DPM-Solver++ \cite{dpmsolver} to denoise the diffusion model.
We train the SVC model for 600K steps with a batch size of 128 and an initial learning rate of $3 \times 10^{-4}$. We use the learning rate schedule with a decay factor of $0.999^{1/8}$ updated at each epoch. We use the AdamW optimizer, with $\beta_1 = 0.8$ and $\beta_2 = 0.99$. We use automatic mixed precision with FP16 for efficient training. 
We utilize pretrained BigVGAN\footnote{\url{https://huggingface.co/nvidia/bigvgan_v2_24khz_100band_256x}} \cite{bigvgan} as a vocoder.

The style encoder of the pitch style converter consists of 4 layers of feed-forward transformer (FFT) blocks \cite{fastspeech}, following the multi-layer perceptron (MLP) layer. The FFT blocks are set an encoder dropout rate of 0.2, a convolution kernel size of 31, and other hyperparameters are same as \cite{fastspeech}. The style embedding dimension is set to 256. For the MPD, we follow the settings in \cite{vits}.
We train the converter for 200K steps with a batch size of 64. We set an initial learning rate of $1 \times 10^{-4}$ for the converter and $1 \times 10^{-5}$ for the discriminator. The optimizer settings are the same as those in the SVC model.

\begin{figure}
  \centering
  \includegraphics[width=\linewidth]{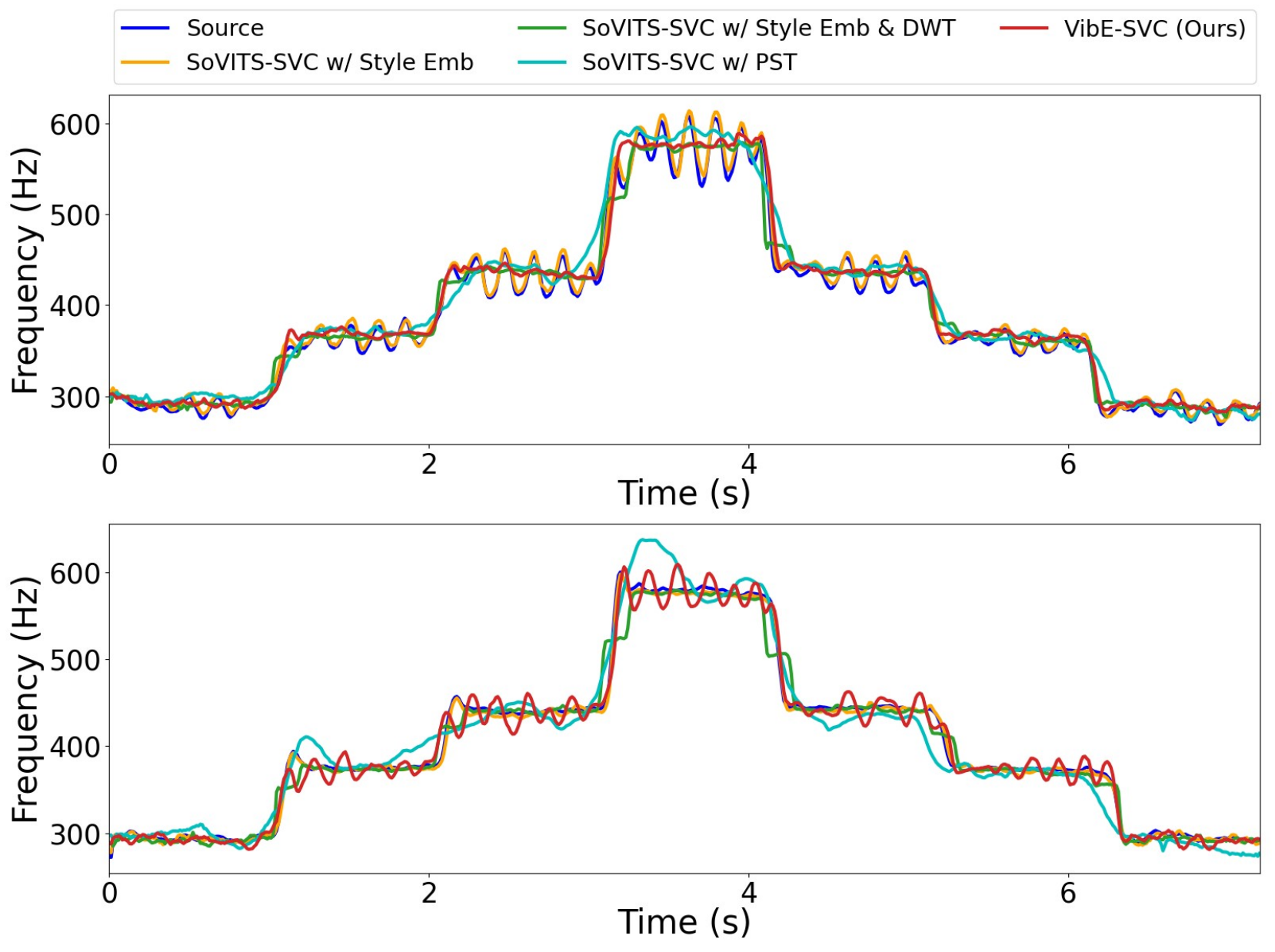}
  \vspace{-0.6cm}
  \caption{
  Comparison of F0 contours: vibrato-to-straight conversion (top) and straight-to-vibrato conversion (bottom).
  }
  \vspace{-0.4cm}
  \label{fig:conversion_visualization}
\end{figure}

\subsection{Baseline models}
We construct baseline models with style embedding based on the SoVITS framework. The first baseline, SoVITS with style embedding, incorporates a style embedding directly on the diffusion decoder, enabling the model to learn style information. The hidden dimension is set to 256. The second baseline, SoVITS with style embedding and DWT, shares the same architecture but replaces the source F0 contour with its low-frequency contour. This design allows the model to capture style information by reconstructing the source F0 contour. We include the SoVITS with F0 model, employed in the performance style transfer (PST) \cite{PST}. To predict singing styles, we replace the speaker embedding with a style embedding.

\subsection{Evaluation metrics}
For objective evaluation, we evaluate speaker similarity using speaker encoder cosine similarity (SECS) with Resemblyzer\footnote{\url{https://github.com/resemble-ai/Resemblyzer}}. To evaluate style transfer quality, we use MERT \cite{mert}, a self-supervised model designed for music information retrieval tasks. We use a pretrained checkpoint\footnote{\url{https://huggingface.co/m-a-p/MERT-v1-330M}} and fine-tune an additional MLP layer to classify singing styles as either straight or vibrato.
For subjective evaluation, we conduct 5-point mean opinion score (MOS) and 4-point similarity mean opinion score (SMOS) tests via Amazon MTurk where at least 20 participants evaluate the naturalness and speaker similarity of 50 samples per model.

\begin{figure}
  \centering
  \begin{subfigure}[t]{0.45\linewidth}
  \includegraphics[width=\linewidth]{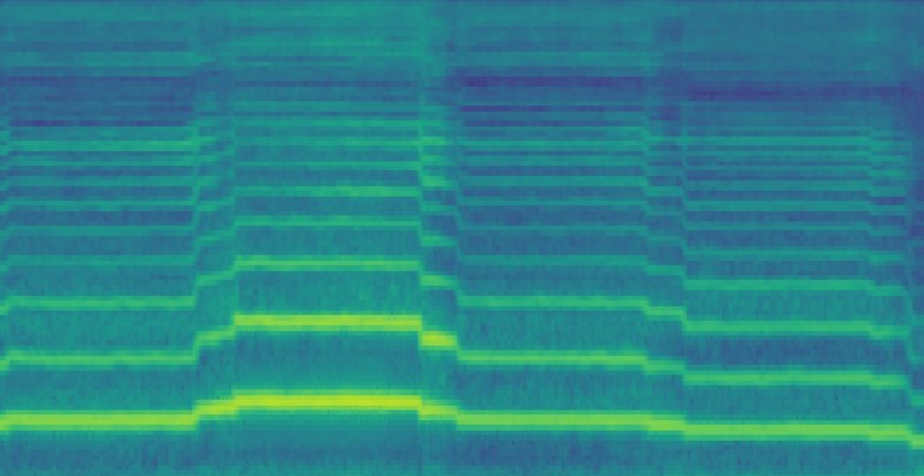}
  \vspace{-0.4cm}
  \caption{Scaling 0.1}
  \label{fig:scaling_01}
  \end{subfigure}
  \begin{subfigure}[t]{0.45\linewidth}
  \includegraphics[width=\linewidth]{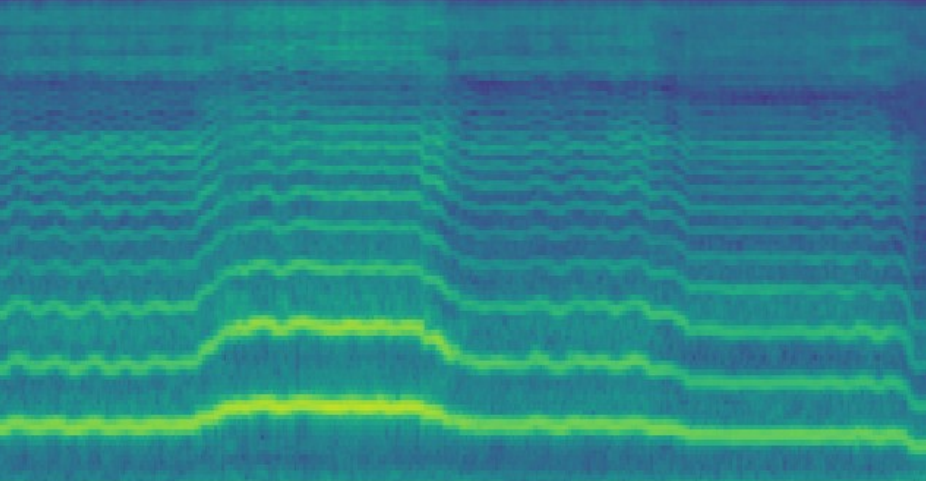}
  \vspace{-0.4cm}
  \caption{Scaling 0.5}
  \label{fig:scaling_05}
  \end{subfigure}
  \begin{subfigure}[t]{0.45\linewidth}
  \includegraphics[width=\linewidth]{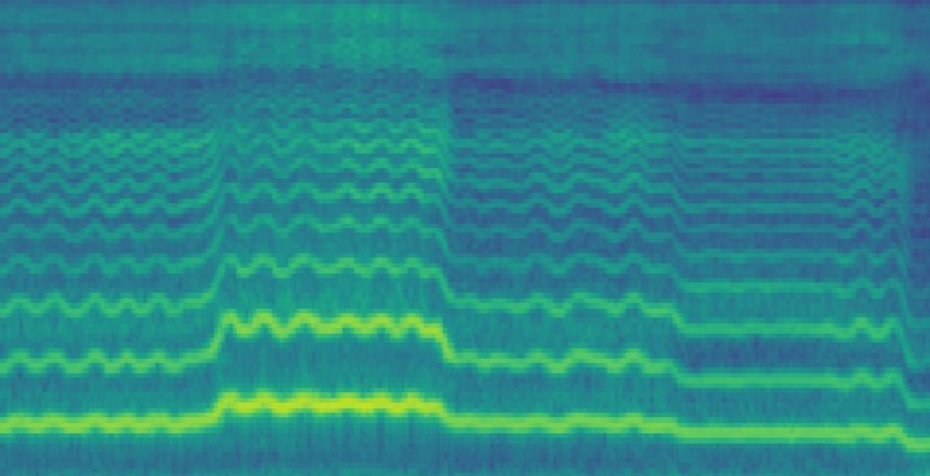}
  \vspace{-0.4cm}
  \caption{Scaling 1.0}
  \label{fig:scaling_10}
  \end{subfigure}
  \begin{subfigure}[t]{0.45\linewidth}
  \includegraphics[width=\linewidth, height = 1.85cm]{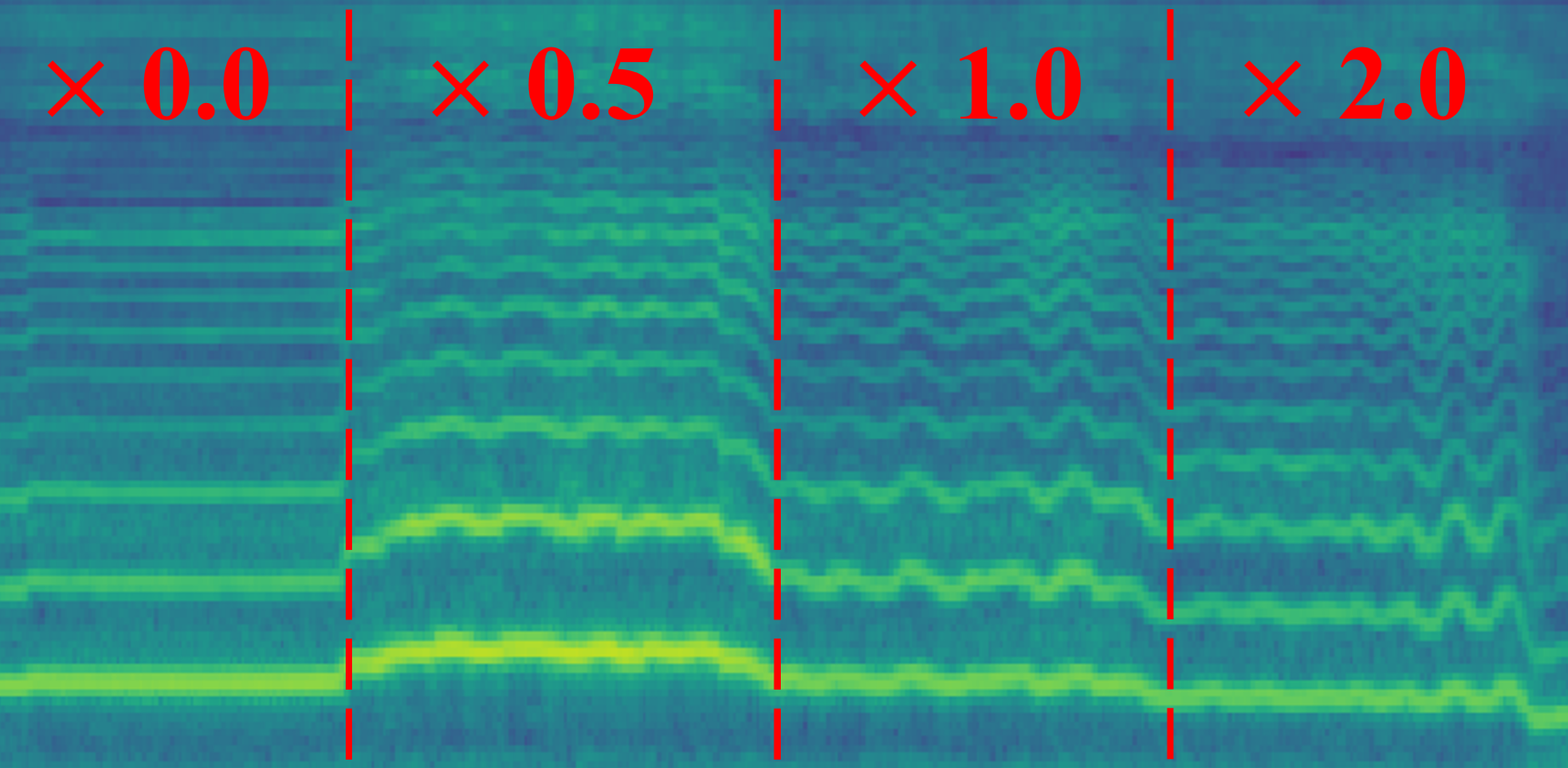}
  \vspace{-0.4cm}
  \caption{Frame-level scaling}
  \label{fig:scaling_frame_level}
  \end{subfigure}
  \vspace{-0.2cm}
  \caption{(a), (b), and (c) show global-level vibrato scaling; (d) shows frame-level vibrato scaling.}
    \vspace{-0.3cm}
  \label{fig:vibrato scaling}
\end{figure} 

\section{Results}

\subsection{Style transfer}
\subsubsection{Objective evaluation}
We conduct two experiments to evaluate style transfer performance. The first focuses on style-only conversion, aiming to assess style transfer without considering speaker conversion quality. The second experiment evaluates pitch style transfer performance in a joint timbre and style conversion setting to assess the model's adaptation ability for the SVC task. As shown in Table \ref{tab:main_table}, our method outperforms baseline models in SECS and style accuracy, demonstrating superior timbre and style conversion in both experiments. 
The experimental results confirm that our model converts singing styles more effectively than baseline model while preserving speaker similarity.
The consistent style accuracy in both experiments shows that our model preserves style information after conversion.

\subsubsection{Subjective evaluation}
As shown in Table \ref{tab:main_table}, our method achieves comparable results in subjective evaluations in terms of speaker similarity and naturalness.
In both experiments, our model shows slight difference results with 95\% confidence intervals in naturalness and speaker similarity across baseline models.
Figure \ref{fig:mos_acc_corr} shows the trade-off between naturalness and style transfer performance among baseline models in both experiments. SoVITS with style embedding achieves the highest naturalness but has the lowest style accuracy, whereas other baseline models exhibit higher style transfer performance at the cost of reduced naturalness. Although our model shows slightly lower naturalness in style-only conversion and slightly lower speaker similarity in timbre and style conversion, the model shows highest style accuracy, balancing effectively this trade-off. These results indicate that the model successfully converts the styles without affecting naturalness and speaker similarity performance.

\subsubsection{Comparison of the F0 contour}
Figure \ref{fig:conversion_visualization} compares the converted F0 contours for vibrato-to-straight and straight-to-vibrato conversions. In the vibrato-to-straight conversion, the baseline models fail to fully remove vibrato or produce unnatural contours. In contrast, our method generates a relatively flat and natural F0 contour, effectively removing the vibrato.
For the straight-to-vibrato conversion, the baseline models struggle to produce noticeable vibrato patterns. However, VibE-SVC successfully generates a clear and natural vibrato, demonstrating superior style conversion performance.

\subsection{Vibrato scaling control}
We conduct an experiment to evaluate the fine-grained vibrato control capability of VibE-SVC. As shown in Figure \ref{fig:vibrato scaling}, the model adjusts vibrato extent by multiplying the high-frequency F0 contour with a constant value before adding the low-frequency F0 contour during inference. Frame-level vibrato scaling is also possible by applying scaling at specific target indices.
To verify that the high-frequency F0 contour contains style information, we perform straight-to-vibrato conversion experiments with different scaling factors using the fine-tuned MERT \cite{mert}. As shown in Table \ref{tab:vibrato_scaling_acc}, the decrease in style accuracy with scaling factors confirms our assumption. VibE-SVC is also able to emphasize vibrato by scaling the high-frequency F0 contour to 2, as shown in Figure \ref{fig:scaling_frame_level}.

\begin{table}[t]
  \caption{The style accuracy with various vibrato scaling.}
  \vspace{-0.25cm}
  \label{tab:vibrato_scaling_acc}
  \centering
  \resizebox{1.00\linewidth}{!}{
  \begin{tabular}{c|cccccc}
    \toprule
    \multicolumn{1}{c|}{\textbf{Scaling Factor}} & 
    \multicolumn{1}{c}{\textbf{0.1}} &
    \multicolumn{1}{c}{\textbf{0.3}} &
    \multicolumn{1}{c}{\textbf{0.5}} &
    \multicolumn{1}{c}{\textbf{0.7}} &
    \multicolumn{1}{c}{\textbf{1.0}} &
    \multicolumn{1}{c}{\textbf{2.0}}\\
    \midrule
    
    \textbf{Style Acc} 
    &\text{0.066} &\text{0.093} &\text{0.217}  
    &\text{0.421} &\text{0.690} &\text{0.928}\\
    
    \bottomrule
  \end{tabular}
  }
  \vspace{-0.1cm}
\end{table}

\begin{table}[t]
  \caption{Ablation study for disentanglement level of DWT.}
  \vspace{-0.25cm}
  \label{tab:dwt_level_ablation}
  \centering
  \begin{tabular}{c|cc|ccc }
    \toprule
    \multicolumn{1}{c|}{\textbf{Level}} & 
    \multicolumn{1}{c}{\textbf{MOS}} & 
    \multicolumn{1}{c|}{\textbf{SMOS}} & 
    \multicolumn{1}{c}{\textbf{SECS}} & 
    \multicolumn{1}{c}{\textbf{Acc}} \\
    \midrule

    \textbf{3}
    &\textbf{3.905 $\pm$ 0.09} &\textbf{2.797 $\pm$ 0.09}
    &\textbf{0.775} &\text{0.163}\\
    
    \textbf{4}
    &\text{3.777 $\pm$ 0.10} &\text{2.796 $\pm$ 0.10}
    &\text{0.774} &\textbf{0.694}\\
    
    \textbf{5}
    &\text{3.646 $\pm$ 0.11} &\text{2.792 $\pm$ 0.09} 
    &\text{0.772} &\textbf{0.694}\\
    \bottomrule
  \end{tabular}
  \vspace{-0.1cm}
\end{table}

\subsection{Ablation studies}
We conduct ablation studies to evaluate the effectiveness of each component in VibE-SVC. As shown in Table \ref{tab:main_table}, removing the MPD decreases naturalness and style accuracy, confirming its importance for singing style modeling. Replacing the high-frequency F0 prediction with the source F0 prediction further reduces both metrics, highlighting the effectiveness of the DWT-based disentanglement method.

To analyze the effect of DWT decomposition levels on performance, we conduct additional experiments. Table \ref{tab:dwt_level_ablation} presents the results for different DWT levels. Level 3 shows the lowest style accuracy due to the absence of vibrato information in high-frequency F0 contour. Although level 5 captures more high-frequency information, it slightly underperforms compared to level 4 because it contains unrelated information beyond vibrato styles. Based on these results, level 4 is chosen as the optimal decomposition level.

\section{Conclusions}
In this paper, we propose a controllable SVC model that designed to adjust singing styles such as vibrato. To disentangle singing style from the F0 contour, we propose a method that decomposes it into low- and high-frequency components using DWT. By predicting the high-frequency F0 contour, the model convert arbitrary inputs into either straight or vibrato styles. Experimental results from both objective and subjective evaluations demonstrate the effectiveness of our approach. 
The current model focuses on controlling vibrato, serving as a starting point for exploring a wider range of singing styles. In future work, we plan to extend the model to capture additional styles related to timbre and aperiodic pitch variations, further improving its applicability.

\section{Acknowledgements}
This work was partly supported by the Institute of Information \& Communications Technology Planning \& Evaluation (IITP) grant funded by the Korea government (MSIT) (Artificial Intelligence Graduate School Program (Korea University) (No. RS-2019-II190079), Artificial Intelligence Innovation Hub (No. RS-2021-II212068), AI Technology for Interactive Communication of Language Impaired Individuals (No. RS-2024-00336673), and Artificial Intelligence Star Fellowship Support Program to Nurture the Best Talents (IITP-2025-RS-2025-02304828)).

\bibliographystyle{IEEEtran}
\bibliography{paper_bib}
\end{document}